\documentclass[cameraready]{Interspeech}

\usepackage{url}
\usepackage{wrapfig}
\usepackage{fontawesome5}
\usepackage{multirow}
\title{Mitigating Scoring Errors and Compensating for Nonverbal Subtests \\ in Speech-Based Dementia Assessment}
\author[affiliation={1}]{Franziska}{Braun}
\author[affiliation={1}]{Christopher}{Witzl}
\author[affiliation={2}]{Andreas}{Erzigkeit}
\author[affiliation={3}]{Hartmut}{Lehfeld}
\author[affiliation={3}]{Thomas}{Hillemacher}
\author[affiliation={1}]{Tobias}{Bocklet}
\author[affiliation={1}]{Korbinian}{Riedhammer}

\address{
    $^1$ Technische Hochschule Nürnberg,
    $^2$ Geromed GmbH,
    $^3$ PMU Klinikum Nürnberg, Germany
}

\email{franziska.braun@th-nuernberg.de}

\keywords{dementia screening, pathological speech}

\usepackage{comment}

\begin{document}

\maketitle

\begin{abstract}
Early detection of cognitive impairment relies on neuropsychological tests to minimize subjectivity by assessing multiple cognitive domains. 
Speech-based evaluation can support diagnostics and improve accessibility, but transcription errors and the omission of nonverbal subtests (e.g., motor skills) limit accuracy. 
Beyond conventional test scores, speech-derived features can provide additional insights into cognitive status. 
This study investigates the speech-based evaluation of the German ``Syndrom-Kurz-Test,'' a standardized dementia screening test comprising verbal and motor subtests. 
We train models that integrate transcript-derived scores and Whisper embeddings per verbal subtest to reduce scoring errors. 
To compensate for missing motor subtests, we then leverage these fused representations to approximate expert overall ratings. 
Despite omitting subtests, our models strongly correlate with expert ratings and efficiently and accurately discriminate between cognitive status groups.
\end{abstract}
\section{Introduction}
Gold standard dementia screening relies on neuropsychological tests, which, together with medical biomarkers, reduce subjectivity by quantifying performance across multiple cognitive domains (e.g., memory, language, and motor functioning).
Speech-based assessments offer a non-invasive, cost-effective, and accessible approach that, in addition to automating clinical protocols, can yield dementia-related biomarkers.
Most tests are performed verbally, making them suitable for automated scoring based on speech-to-text transcription. 
Moreover, paralinguistic and linguistic speech features in test responses can provide diagnostically relevant information that goes beyond conventional aggregate test scores.
However, the development of speech-based assistive tools faces two major challenges: \textbf{(1)} the target speech is especially susceptible to transcription errors, owing to the increased prevalence of dialectal, pathological, and atypical speech patterns (e.g., structured test responses); and \textbf{(2)} not all assessment objectives can be covered by speech-based methods, as some tasks rely on motor skills rather than spoken responses.
This study addresses both challenges in an end-to-end, speech-based approach to automating the evaluation of the German Syndrom-Kurz-Test (SKT).
The SKT is a standardized neuropsychological test battery, comprising verbal subtests such as naming, recalling, reading, and counting, as well as motor subtests such as sorting and returning game tokens.  
Speech data was collected in the context of routine clinical practice, in which the SKT is part of a face-to-face dementia screening procedure.  
During administration, experts manually scored subtests by measuring processing time or counting missing test responses while maintaining continuous patient interaction and observation.
Scores are normalized and aggregated into the SKT total score, which allows interpretation on a six-point ordinal scale from no cognitive impairment to very severe dementia.
To support these administratively intensive processes, the main contributions of this study are as follows:  
\textbf{(1)} We establish a rule-based baseline to calculate subtest scores from SKT responses using Whisper transcripts and quantify the deviation from expert-derived scores.
\textbf{(2)} To reduce transcription-induced scoring errors, our \textit{deep correction} models combine rule-based scores with their underlying Whisper embeddings to predict refined, expert-approximate subtest scores.  
\textbf{(3)} Given that motor subtests cannot be evaluated via speech and others only to a limited extent, our \textit{deep compensation} models combine the \textit{deep correction} models from the available subtests to predict expert-approximate SKT total scores.  
\textbf{(4)} We explore the optimal subtest sequence for speech-based assessment that maximizes overall dementia classification accuracy and efficiency.
\section{Related Work}
Research in speech-based dementia assessment mainly focuses on two principal approaches: (1) the feature extraction from elicited speech (e.g., picture descriptions) to predict and classify cognitive impairment utilizing established assessment scales \cite{adress20,adresso21,madress22,taukadial24,braun_classifying_2023} such as the MMSE \cite{mmse75} and MoCA \cite{moca_2005}; and (2) the application of speech processing techniques to automate standardized neuropsychological tests (e.g., Verbal Fluency, Boston Naming) \cite{troeger18,koenig18,kim19,kwon21,lofgren_breaking_2022,braun22_interspeech,braun_GoingCookieTheft_2022}.
Extensive studies on ADReSS \cite{adress20}, ADReSSo \cite{adresso21}, MADReSS \cite{madress22}, and TAUKADIAL \cite{taukadial24} challenges have highlighted that while distinguishing between speech from dementia and healthy control groups is straightforward, early detection in terms of mild cognitive impairment (MCI) remains a major challenge.
Most studies focus on isolated speech tasks from test batteries (e.g., picture descriptions from BDAE \cite{borod80}), even though these batteries are designed to jointly assess multiple cognitive domains for a holistic diagnostic profile, so single-task analyses may reduce sensitivity for early detection.
The SKT \cite{erzigkeit77, erzigkeit15} battery differentiates cognitive impairment on a six-point scale based on attention and memory profiles drawn from established neuropsychological test concepts.
Prior work has confirmed that the automated evaluation of the SKT is feasible, but the atypical speech content of certain subtests (e.g., counting) poses challenges for both traditional (DNN-HMM) and neural (OWSM, Whisper) ASR-systems \cite{braun22_interspeech,braun25_interspeech}.
The study also showed that the speech from SKT subtests contains markers for cognitive impairment beyond conventional test scores 
\cite{braun_GoingCookieTheft_2022}.
Whisper models \cite{whisper} provide SOTA German ASR and embeddings that have been successfully applied to classify pathological speech, such as dysarthria \cite{rathod23_interspeech}, stuttering \cite{changawala24_interspeech,wagner24b_interspeech}, and dementia \cite {brainsci14121292}.
We extend existing studies by leveraging the ASR and embedding capabilities of Whisper models to improve the automated evaluation of the SKT, enabling accurate and efficient early dementia detection.
\section{Data}\label{sc:data}
%
%
We use a subset of the corpus introduced in \cite{braun22_interspeech}, which comprises 158 German-speaking subjects (63 men, 95 women) aged between 49 and 89 years ($\mu = 73.69 \pm $9.02).
Demographic data for the diagnostic groups of no cognitive impairment (NCI), mild cognitive impairment (MCI), and dementia (DEM) is listed in \cite[Table~1]{braun25_interspeech}.
All tests and recordings were conducted during routine clinical practice as part of a face-to-face dementia screening procedure that included history taking, cognitive testing, and questionnaires.
Speakers wore surgical masks, and some exhibited strong local dialects, which can present challenges for ASR systems.
Per subject, we use speech samples from four subtests measuring \textbf{attention performance} (SKT1 = naming objects, SKT3 = reading numbers, SKT6 = counting symbols, SKT7 = reading letters/interference test) and three subtests measuring \textbf {memory performance} (SKT2 = recalling objects immediately, SKT8 = recalling objects delayed, SKT9 = recognizing objects from distractors \cite{erzigkeit15}.
We exclude the tasks of ordering (SKT4) and returning numbers (SKT5), which require actions instead of speech.
All tests were administered and manually evaluated by experts in strict accordance with the SKT guidelines \cite{erzigkeit15}.
Detailed descriptions of the SKT tasks and evaluation can be found in \cite{erzigkeit15, braun22_interspeech}.
\section{Method}
\subsection{Whisper Transcripts and Embeddings}\label{sc:transcripts}
\begin{table}[th]
\centering
\caption{\label{tab:wer}Syndrom-Kurz-Test (SKT) verbal (\faCommentDots[regular]) and motor (\faHandPaper[regular]) subtests 1--9: WER (in \%) for the whisper models (small and large-v3), and Pearson Correlation ($r$) with the SKT total score.}
\begin{tabular}{@{}@{}l||l|l|l|l|l|l@{}}
\toprule
\multicolumn{1}{c||}{\textbf{SKT}} & \multicolumn{1}{c|}{\textbf{task}} & \multicolumn{1}{c|}{\textbf{function}} & \multicolumn{1}{c|}{\textbf{scoring}} & \multicolumn{1}{c|}{\textbf{small}} & \multicolumn{1}{c|}{\textbf{large}} & \multicolumn{1}{c}{\textbf{$r$}} \\
\midrule
\faCommentDots[regular] \textbf{1} & naming & attention & 0-60 sec & 45.3 & 27.5 & .47 \\
\faCommentDots[regular] \textbf{2} & recall & memory & 0-12 items & 53.0 & 38.4 & .61 \\
\faCommentDots[regular] \textbf{3} & reading & attention & 0-60 sec & 36.3 & 18.7 & .55 \\
\faHandPaper[regular] \textbf{4} & sorting & attention & 0-60 sec & -  & -  & .79 \\
\faHandPaper[regular] \textbf{5} & returning & attention & 0-60 sec & -  & -  & .76 \\
\faCommentDots[regular] \textbf{6} & counting & attention & 0-60 sec & 103.0 & 123.2 & .69 \\
\faCommentDots[regular] \textbf{7} & interference & attention & 0-60 sec & 107.0 & 77.0 & .76 \\
\faCommentDots[regular] \textbf{8} & recall & memory & 0-12 items & 53.0 & 42.5 & .64 \\
\faCommentDots[regular] \textbf{9} & recognizing & memory & 0-12 items & 47.6 & 31.5 & .59 \\ 
\bottomrule
\end{tabular}
\end{table}
For automatic transcription and feature extraction, we use OpenAI's encoder-decoder-based model Whisper; Model weights are open source and can be accessed online (huggingface).
We compare \texttt{whisper-small} and \texttt{whisper-large-v3}, which is one of the SOTA ASR models for the German language and has already proven to be the most suitable for our purposes in previous work compared to other architectures (e.g., DNN-HMM, OWSM, Parakeet) \cite{braun22_interspeech,braun25_interspeech,braun26_lrec}.
We transcribe to German using beam search (beam\_size=5) with early stopping and return word-level timestamps. 
The word error rates (WERs) for the subtests and their respective transcripts are given in \tablename~\ref{tab:wer}. 
WERs above 100\% result from hallucinated insertions in the transcripts caused by long pauses when counting (SKT6) and atypical sequences such as ``ABBABA...'' in the interference test (SKT7).
A parameter to limit N-gram repetition is used during decoding for these subtests, which partly reduces these effects.
The complete normalized wave files (duration = 8--155 s) are processed.
During the transcription process, we also extract the encoder and decoder embeddings that underlie the model's ability to process and generate text representations from speech.
These embeddings effectively bridge the gap between audio and text representations and capture complex acoustic and linguistic features of the speech data.
We obtain the final embeddings from the last model layer without pooling, which corresponds to 768- and 1280-dimensional feature vectors per frame or token in the input sequence for \texttt{whisper-small} and \texttt{whisper-large-v3}.
\subsection{Rule-based Scoring}\label{sc:automatic_scores}
For automated scoring, we only consider subtests that can be evaluated using an audio recording and the corresponding transcript. 
Therefore, we exclude the tasks sorting (4) and returning (5) numbers, which are performed by hand.
First, we interpolate the token timestamps to obtain fine-grained word-level timestamps. 
The raw subtest scores reflect the processing time (0--60 seconds) in the attention tests and the number of missing objects (0--12) in the memory tests.
For the memory subtests (2, 8, and 9), the score is calculated based on the difference between the recognized objects, including synonyms from a predefined dictionary, and the expected objects.
For the attention subtests (1, 3, 6, and 7), the timestamp of the last word that matches the task-specific expected content (e.g., objects, numbers, letters) represents the score.
Normalized for age and education according to \cite[Sec.~7]{erzigkeit15}, this results in norm scores (from 0--3) for each subtest.
The sum of the norm scores gives the total SKT total score (0--27), categorized into no cognitive impairment (0--4), mild cognitive impairment (5--8), and mild (9--13), moderate (14--18), severe (19--23), and very severe (24--27) dementia.
\section{Experiments}
\begin{figure*}
    \centering
    \includegraphics[width=0.9\linewidth]{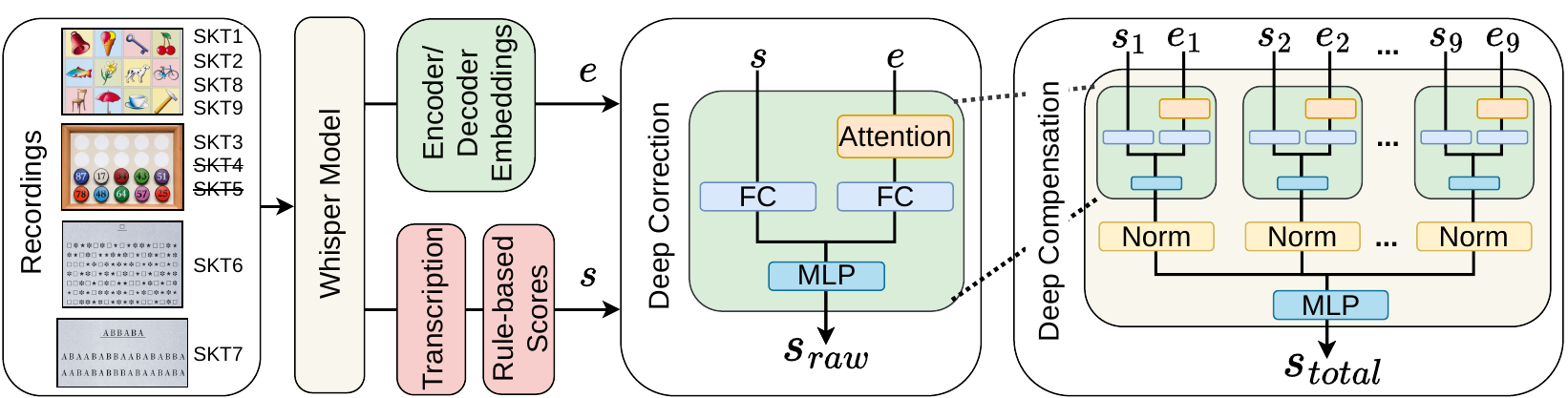}
    \caption{Speech-based Dementia Assessment using the Syndrom-Kurz-Test (SKT).}
    \label{fig:model}
\end{figure*}
All experiments were conducted using \texttt{whisper-small} and \texttt{whisper-large-v3} transcripts and their underlying encoder and decoder embeddings.
To assess model performance, we employ stratified five-fold cross-validation, partitioning the dataset into five speaker-distinct training (80\%) and test (20\%) sets.
We use Root Mean Square Error (RMSE) to quantify the regression models' performance, which measures the average deviation between predicted and actual values.
In addition, we report the Pearson correlation between the expert and predicted scores.
We use MSE loss function and early stopping (patience=5, delta=0) during training.
For consistency and comparability, all experiments are conducted in a fixed hyper-parameter setting (optimizer=Adam, batch\_size=8, max\_epochs=100, activation\_function=ReLU) without tuning.
%
\subsection{Mitigating Transcription-Related Scoring Errors}\label{sc:deepcorrection}
To mitigate errors resulting from transcription and scoring process, we train regression models that take as input the rule-based (RB) scores calculated from transcripts in combination with the encoder (ENC) and decoder (DEC) embeddings underlying these transcripts.
This approach aims to train models to correct scores based on latent speech representations; it is referred to as \textit{deep correction} in the following.
The model architecture is shown in \figurename~\ref{fig:model}.
The inputs consist of RB scores $s$ and embedding vectors $e$. 
Before fusion, $e$ is first passed to a single-head self-attention block, followed by mean pooling and layer normalization. 
The resulting representations of $s$ (dim=1) and $e$ (dim=768/1024) are fed into separate fully connected (FC) layers (output\_dim=256) for equal dimensionality and balanced contributions during training.
Subsequently, $s$ and $e$ are concatenated and passed to a 2-layer MLP (hidden\_dim=64), which outputs the corrected raw scores $s_{raw}$.
We train models (lr=1e-3) for each subtest, using the expert-assigned raw scores as ground truth (GT) labels. 
In addition, we train models (lr=1e-5) that only receive the encoder or decoder embeddings without RB inputs to investigate whether the scores can also be learned from embeddings alone.
\subsection{Compensating for Nonverbal Subtests}\label{sc:deepcompensation}
In order to enable efficient and accurate speech-based dementia assessment without motor subtests, we are investigating the subtest sequence that maximizes the overall test result (i.e., the SKT total score) in minimum steps.
We achieve this by sequentially ordering the subtests and discontinuing as soon as the probability of a particular diagnosis exceeds a predefined decision threshold (correlation of 0.9). 
Therefore, the raw scores from Sec.~\ref{sc:deepcorrection} are converted to norm scores, which in sum give the SKT total score. 
The correlation with the expert norm scores is given in parentheses in \tablename~\ref{tab:pearson}.
We then iteratively calculate the cumulative sum of the subtests' norm scores and compute the correlation ($r_{total}$) with the SKT total score. 
The best order is obtained by adding the respective subtest that maximizes $r_{total}$ at each time step $t$.

In an improved setting, we train regression models to approximate the SKT total scores by iteratively adding the pre-trained subtest models from Sec.~\ref{sc:deepcorrection}.
This approach aims to train models to compensate for missing (nonverbal) subtests in the overall test results; it is referred to as \textit{deep compensation} in the following.
The model architecture is shown in \figurename~\ref{fig:model}.
The model inputs are RB scores $s_{1-9}$ and embeddings $e_{1-9}$ of the added subtests (1--9) at time step $t$, which are fed to the (unfreezed) \textit{deep correction} models to optimize raw scores.
The outputs are normalized, concatenated and passed to an MLP to optimize the SKT total score $s_{total}$.
We train models (lr=1e-2) for each time step, using the expert-assigned SKT total scores as ground truth labels (GT).
\section{Results}
%
\subsection{Mitigating Transcription-Related Scoring Errors}\label{sc:deepcorrection_results}
\begin{table}[th]
\centering
\caption{\label{tab:rmse_errors}Average \textbf{RMSE and STD} for SKT \textbf{raw scores} from models using encoder (ENC) or decoder (DEC) embeddings, and fusion with rule-based scoring (RB).}
\begin{tabular}{@{}c|c||c|c|c|c|c@{}}
\toprule
 & \textbf{SKT} & \textbf{RB} & \textbf{ENC} & \textbf{DEC} & \textbf{RB+ENC} & \textbf{RB+DEC} \\ 
\midrule
\multirow{7}{*}{\rotatebox{90}{\textbf{whisper-small}}} 
 & \textbf{1} & 3.88 & \textbf{2.43$\pm$0.74} & 3.47$\pm$0.37 & 3.38$\pm$1.43 & 3.66$\pm$1.39 \\
 & \textbf{2} & 1.61 & 1.68$\pm$0.28 & 1.62$\pm$0.15 & 1.13$\pm$0.15 & \textbf{1.11$\pm$0.16} \\
 & \textbf{3} & 1.56 & 1.62$\pm$0.82 & 3.02$\pm$2.51 & 1.58$\pm$0.66 & \textbf{1.52$\pm$0.68} \\
 & \textbf{6} & 12.93 & \textbf{3.68$\pm$1.31} & 6.00$\pm$1.34 & 4.67$\pm$1.39 & 5.57$\pm$1.55 \\
 & \textbf{7} & 5.53 & \textbf{3.82$\pm$0.72} & 6.80$\pm$1.41 & 3.87$\pm$0.43 & 4.16$\pm$0.76 \\
 & \textbf{8} & 1.41 & 1.96$\pm$0.35 & 1.78$\pm$0.10 & \textbf{1.07$\pm$0.22} & 1.09$\pm$0.24 \\
 & \textbf{9} & 3.52 & 2.24$\pm$0.50 & 2.34$\pm$0.43 & \textbf{2.04$\pm$0.40} & 2.11$\pm$0.37 \\
\midrule
\multirow{7}{*}{\rotatebox{90}{\textbf{whisper-large-v3}}} 
 & \textbf{1} & 3.24 & \textbf{2.66$\pm$0.58} & 4.28$\pm$0.90 & 2.90$\pm$1.48 & 2.66$\pm$1.12 \\
 & \textbf{2} & 0.84 & 1.70$\pm$0.19 & 1.52$\pm$0.21 & 0.76$\pm$0.10 & \textbf{0.74$\pm$0.13} \\
 & \textbf{3} & 1.21 & 1.51$\pm$1.42 & 2.23$\pm$1.62 & \textbf{1.15$\pm$0.41} & \textbf{1.15$\pm$0.43} \\
 & \textbf{6} & 8.12 & 5.16$\pm$1.13 & 6.31$\pm$1.48 & \textbf{3.70$\pm$1.24} & 4.43$\pm$2.11 \\
 & \textbf{7} & 5.94 & \textbf{4.09$\pm$0.38} & 6.78$\pm$0.47 & 4.25$\pm$1.47 & 4.68$\pm$1.92 \\
 & \textbf{8} & 0.82 & 1.96$\pm$0.45 & 1.59$\pm$0.16 & \textbf{0.75$\pm$0.19} & 0.76$\pm$0.13 \\
 & \textbf{9} & 2.12 & 2.45$\pm$0.56 & 2.27$\pm$0.25 & \textbf{1.63$\pm$0.32} & 1.67$\pm$0.37 \\
\bottomrule
\end{tabular}
\end{table}
\begin{table}[th]
\centering
\caption{\label{tab:pearson}\textbf{Pearson Correlation} with experts for SKT \textbf{raw} and \textbf{(norm) scores} from models using encoder (ENC) or decoder (DEC) embeddings, and fusion with rule-based scoring (RB).}
\begin{tabular}{@{}c|c||c|c|c|c|c@{}}
\toprule
 & \textbf{SKT} & \textbf{RB} & \textbf{ENC} & \textbf{DEC} & \textbf{RB+ENC} & \textbf{RB+DEC} \\ 
\midrule
\multirow{7}{*}{\rotatebox{90}{\textbf{whisper-small}}}
 & \textbf{1} & .93 (.94) & \textbf{.97 (.94)} & .93 (.90) & .93 (.93) & .92 (.94) \\
 & \textbf{2} & \textbf{.79 (.70)} & .44 (.19) & .54 (.44) & \textbf{.79 (.69)} & \textbf{.79 (.68)} \\
 & \textbf{3} & \textbf{.96 (.85)} & .95 (.89) & .78 (.81) & \textbf{.96 (.85)} & \textbf{.96 (.85)} \\
 & \textbf{6} & .59 (.69) & \textbf{.94 (.89)} & .83 (.76) & .89 (.86) & .86 (.81) \\
 & \textbf{7} & .92 (.95) & \textbf{.95 (.91)} & .83 (.76) & \textbf{.95 (.91)} & \textbf{.95 (.93)} \\
 & \textbf{8} & .90 (.86) & .70 (.68) & .76 (.73) & \textbf{.91 (.87)} & \textbf{.91 (.87)} \\
 & \textbf{9} & .64 (.57) & .60 (.51) & .57 (.55) & .68 (.58) & \textbf{.74 (.66)} \\
\midrule
\multirow{7}{*}{\rotatebox{90}{\textbf{whisper-large-v3}}}
 & \textbf{1} & \textbf{.96 (.98)} & \textbf{.96 (.94)} & .89 (.88) & .94 (.90) & .95 (.94) \\
 & \textbf{2} & \textbf{.92 (.87)} & .42 (.30) & .54 (.41) & .89 (.83) & .90 (.83) \\
 & \textbf{3} & \textbf{.98 (.94)} & .94 (.92) & .92 (.85) & \textbf{.98 (.94)} & \textbf{.98 (.94)} \\
 & \textbf{6} & .78 (.85) & .89 (.85) & .81 (.81) & \textbf{.94 (.90)} & .90 (.84) \\
 & \textbf{7} & .90 (.93) & \textbf{.95 (.89)} & .84 (.75) & .94 (.90) & .92 (.90) \\
 & \textbf{8} & \textbf{.96 (.93)} & .70 (.65) & .81 (.79) & .95 (.93) & .95 (.93) \\
 & \textbf{9} & .80 (.76) & .51 (.49) & .60 (.57) & \textbf{.82 (.72)} & .79 (.71) \\
\bottomrule
\end{tabular}
\end{table}
The results for the \textit{deep correction} models (Sec.~\ref{sc:deepcorrection}) are given as RMSE in \tablename~\ref{tab:rmse_errors} and Pearson correlation \tablename~\ref{tab:pearson}.
The predicted values for the attention subtests (1, 3, 6, 7) vary from 0--60, while the memory subtests (2, 8, 9) vary from 0--12 (cf. \tablename~\ref{tab:wer}).
For all subtests, the RB+ENC and RB+DEC predicted scores show strong to very strong correlations with expert scores, suggesting that RB scoring errors can be effectively mitigated by incorporating embedding information.
A kind of ``component balancing'' can be observed, so that in the ASR-demanding subtests with more errors occurring in the RB component, models can correct the results with the additional embedding information, increasing correlations by up to 0.35 (task 6). 
Meanwhile, in the simpler ASR tasks (2, 3, 8), the model results appear to rely on the already robust RB assessment.
Specifically, subtests 1, 6, and 7 benefit from ENC, suggesting that encoder information helps with, for example, hallucination-related errors that occur when subjects count silently or read atypical letter sequences.
In the attention tests, models incorporating ENC show accurate predictions of processing time and may potentially reduce timestamp errors. 
In the memory tests, the RB approach is particularly advantageous for predicting named objects, with the exception of subtest 9, where the embedding space may help to reduce task-specific errors, such as when subjects point to objects instead of naming them or make use of negation (e.g., ``There was no dog there'').
\subsection{Compensating for Nonverbal Subtests}
\begin{table}[th]
\centering
\caption{\label{tab:rmse_missing}Average \textbf{RMSE and STD} for SKT \textbf{total scores} at time step $t$ from models using encoder embeddings (ENC) or decoder (DEC) embeddings, and fusion with rule-based scoring (RB).}
\begin{tabular}{@{}c|c||c|c|c|c@{}}
\toprule
 & \textbf{$t$} & \textbf{RB} & \textbf{ENC deep} & \textbf{RB+ENC deep} & \textbf{RB+DEC deep} \\
\midrule
\multirow{7}{*}{\rotatebox{90}{\textbf{whisper-small}}}
 & \textbf{0} & 6.93 & 3.42$\pm$0.68 & \textbf{3.38$\pm$0.70} & 3.40$\pm$0.63 \\
 & \textbf{1} & 5.72 & \textbf{2.69$\pm$0.36} & 2.72$\pm$0.21 & 2.67$\pm$0.21 \\
 & \textbf{2} & 4.67 & \textbf{2.22$\pm$0.31} & 2.24$\pm$0.20 & \textbf{2.22$\pm$0.38} \\
 & \textbf{3} & 3.75 & 2.05$\pm$0.34 & \textbf{2.03$\pm$0.12} & 2.17$\pm$0.15 \\
 & \textbf{4} & 3.25 & 2.03$\pm$0.17 & \textbf{1.98$\pm$0.17} & 2.06$\pm$0.13 \\
 & \textbf{5} & 2.80 & 1.96$\pm$0.29 & \textbf{1.94$\pm$0.24} & 1.99$\pm$0.12 \\
 & \textbf{6} & 2.58 & \textbf{1.91$\pm$0.32} & 1.93$\pm$0.15 & 1.97$\pm$0.13 \\
\midrule
\multirow{7}{*}{\rotatebox{90}{\textbf{whisper-large-v3}}}
 & \textbf{0} & 7.03 & \textbf{3.25$\pm$0.50} & 3.52$\pm$0.67 & 3.34$\pm$0.59 \\
 & \textbf{1} & 5.92 & 2.75$\pm$0.06 & 2.75$\pm$0.41 & \textbf{2.57$\pm$0.28} \\
 & \textbf{2} & 4.89 & 2.29$\pm$0.18 & \textbf{2.21$\pm$0.33} & 2.22$\pm$0.38 \\
 & \textbf{3} & 3.99 & 2.18$\pm$0.20 & 2.04$\pm$0.33 & \textbf{2.01$\pm$0.29} \\
 & \textbf{4} & 3.16 & 2.09$\pm$0.19 & 1.88$\pm$0.38 & \textbf{1.86$\pm$0.38} \\
 & \textbf{5} & 2.54 & 2.09$\pm$0.14 & 1.74$\pm$0.33 & \textbf{1.66$\pm$0.45} \\
 & \textbf{6} & 2.23 & 2.05$\pm$0.21 & \textbf{1.73$\pm$0.32} & \textbf{1.73$\pm$0.40} \\
\bottomrule
\end{tabular}
\end{table}
\begin{figure}[th]
  \centering
  \includegraphics[width=\linewidth]{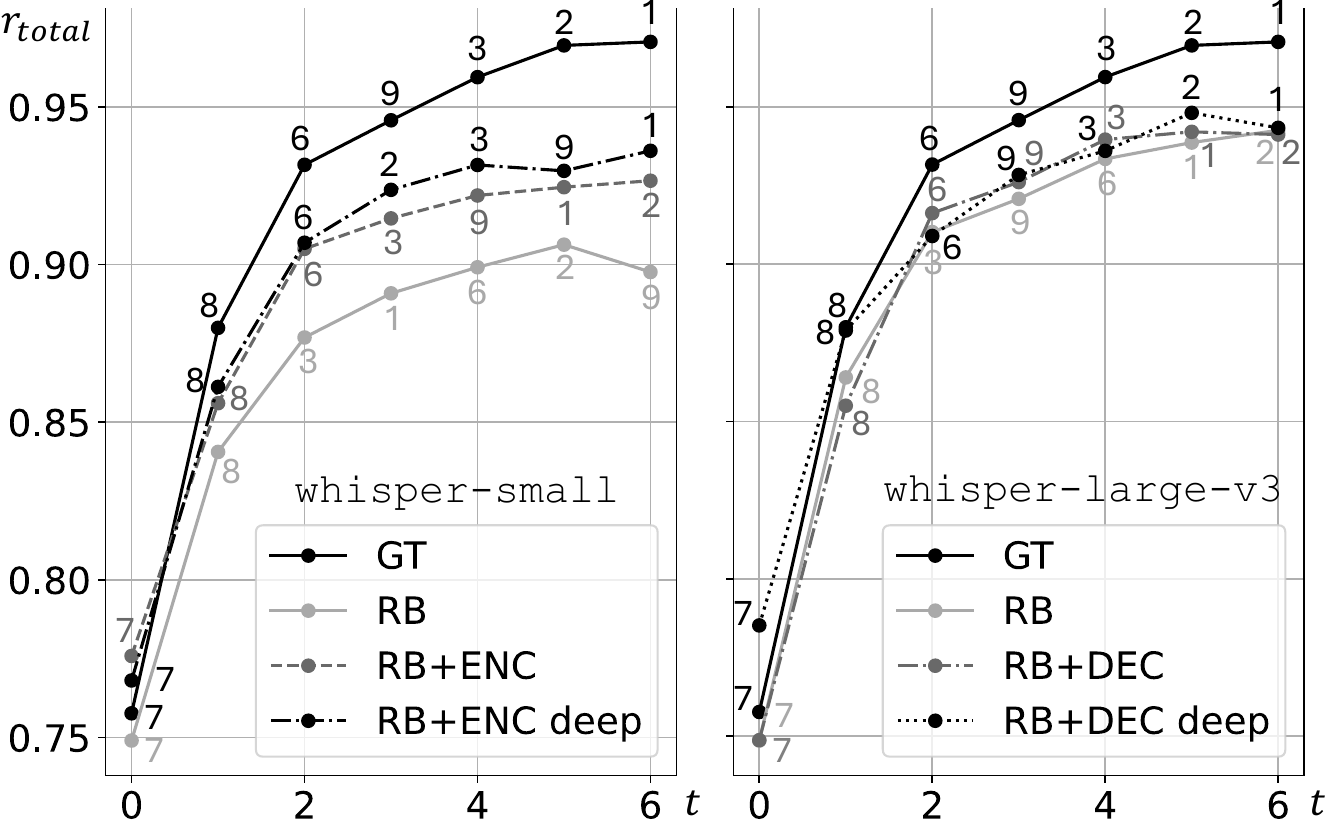}
  \caption{Pearson Correlation with expert SKT total scores ($r_{total}$) by adding subtests (1--9) at time step $t$.}
  \label{fig:plot}
\end{figure}
The correlation with the SKT total score when subtests are added at time step $t$ is shown in \figurename~\ref{fig:plot}; for reasons of clarity and limited space, only the results for the best-performing systems from \tablename~\ref{tab:rmse_missing} are shown.
Despite the exclusion of motor tasks 4 and 5, the results show very strong correlations with the SKT total score when we sequentially process the remaining verbal subtests, reaching up to 0.94 for \texttt{whisper-small} and 0.95 for \texttt{whisper-large}.
\begin{figure}[th]
  \centering
  \includegraphics[width=\linewidth]{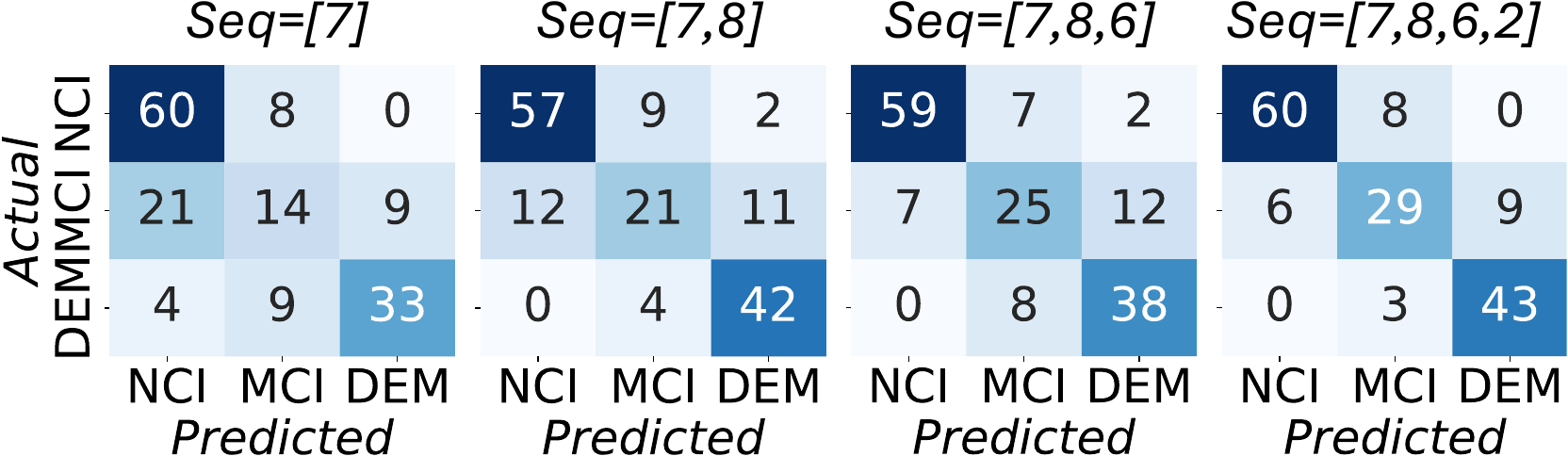}
  \caption{Confusion Matrices (whisper-small) for \textit{deep compensation} (RB+ENC deep) for a given subtest sequence (Seq).}
  \label{fig:conf_small}
\end{figure}

Observing that certain subtests exhibit greater diagnostic sensitivity than others, we are investigating the optimal subtest sequence to achieve accurate dementia classification.
The sequence of subtests 7 (interference), 8 (recall), and 6 (counting) achieves near-perfect correlations above 0.9 across all models, indicating strong diagnostic performance.
\figurename~\ref{fig:conf_small} presents the confusion matrix for diagnostic classification, mapping predicted scores to the corresponding SKT-based classes: no cognitive impairment (NCI), mild cognitive impairment (MCI), and mild to moderate dementia (DEM) (cf. Sec.~\ref{sc:data}). 
The matrix is shown for the \texttt{whisper-small} RB+ENC deep configuration (see Fig.~\ref{fig:plot}). 
Notably, interference (subtest 7) at \textit{$t$=0}, the cognitively most demanding task, effectively discriminates NCI from the impaired groups.
For DEM, adding the recall task (7$\rightarrow$8) at \textit{$t$=1} and for MCI adding the counting task (7$\rightarrow$8$\rightarrow$6) at \textit{$t$=2} further increases the discriminative power.
Incorporating another recall task (7$\rightarrow$8$\rightarrow$6$\rightarrow$2) at \textit{$t$=3} results in robust classification across all diagnostic groups.
In clinical practice, however, recall cannot be administered without the initial stimulus presentation. 
Accordingly, initiating the sequence with the naming task (1$\rightarrow$7$\rightarrow$8$\rightarrow$6$\rightarrow$2) yields an SKT total score correlation of 0.92 (0.93) for \texttt{whisper-small} (\texttt{whisper-large-v3}).
%
%
\section{Conclusion}
We have introduced mitigation methods that utilize both transcripts and their embedded encoder and decoder representations to improve robustness in speech-only dementia assessment.
Using data from routine clinical practice, we observed that our \textit{deep correction} models can mitigate scoring errors for the automated evaluation of the SKT. 
Furthermore, our results suggest that our \textit{deep compensation} models can compensate for the absence of motor subtests and, despite excluding half of the subtests, achieve high correlations with experts overall ratings, enabling efficient and accurate dementia classification. 
While neural ASR models achieve state-of-the-art performance, challenges remain in pathological speech contexts arising from noisy clinical environments with multiple speakers, and atypical or disordered speech.
Transcription errors are particularly critical in clinical and assistive settings, where transcription accuracy directly impacts diagnosis, communication aids, and research, underscoring the need for model adaptations to handle temporal irregularities and non-standard prosody. 
Fine-tuning Whisper on various pathological speech datasets \cite{CrisperWhisper} and hybrid ASR approaches that combine traditional techniques with neural predictions could improve reliability in high-risk contexts. 
Future work should investigate ASR errors across different pathologies to support the development of more specialized systems.
\section{Acknowledgments}
Funded by the Deutsche Forschungsgemeinschaft (DFG) -- Project Number 549142762 -- FIP 160.
\section{Generative AI Use Disclosure}
Generative AI tools were used only for editing and polishing the manuscript; all scientific content, analyses, and conclusions are the responsibility of the authors.
\bibliographystyle{IEEEtran}
\bibliography{template}
\end{document}